\documentclass{article}
\usepackage[T1]{fontenc} 
\usepackage[utf8]{inputenc} 
\usepackage{ismir,amsmath,cite,url}
\usepackage{graphicx}
\usepackage{color}
\def\lbd{}	
\newcommand{\fo}{$f_0$}

\usepackage{lineno}

\title{Encoding Performance Data in MEI with the Automatic Music Performance Analysis and Comparison Toolkit (AMPACT)}

\twoauthors
  {Johanna Devaney} {Brooklyn College and CUNY Graduate Center \\ johanna.devaney@brooklyn.cuny.edu}
  {Cecilia Beauchamp} {The Bronx High School of Science \\ {\tt }}  

\def\authorname{C. Beauchamp and J. Devaney}

\begin{document}

\maketitle
\begin{abstract}
This paper presents a new method of encoding performance data in MEI using the recently added \texttt{<extData>} element. Performance data was extracted using the Automatic Music Performance Analysis and Comparison Toolkit (AMPACT) and encoded as a JSON object within an \texttt{<extData>} element linked to a specific musical note. A set of pop music vocals has was encoded to demonstrate both the range of descriptors that can be encoded in <extData> and how AMPACT can be used for extracting performance data in the absence of a fully specified musical score. 
\end{abstract}
\section{Introduction}\label{sec:introduction}
Typically music performance data is encoded in CSV files, where descriptors (such as tuning) are linked to specific time points in the analyzed audio files\cite{sapp07,hashida2008new,marchini14,dai15}. Some datasets, such as \cite{dai15}, provide musical note information but  note data is typically limited to musical note names with no clear method for including rhythmic or metrical information. Also, while the CSV format is well suited for encoding single-value summaries across the duration of each audio event, it becomes unruly when encoding continuous performance data. This paper describes a new method for encoding both summary and continuous performance descriptors extracted from audio files linked to a symbolic representation of the musical data using the newly introduced \texttt{<extData>} element in MEI. To demonstrate this approach, a small dataset is also presented. The dataset consists of timing-, pitch-, loudness-, and timbre-related descriptors of the vocal tracks in Rihanna's 2016 \emph{Anti} album that were extracted using the Automatic Music Performance Analysis and Comparison Toolkit (AMPACT).\footnote{\url{http://www.ampact.org}}

\section{\texttt{<extData>} in MEI}\label{sec:extData}
MEI \cite{roland02} was primarily designed to encode symbolic musical data. The recent inclusion of the \texttt{<extData>}\footnote{\url{https://music-encoding.org/guidelines/v5/elements/extData.html}}  element in the latest release of MEI\footnote{\url{https://music-encoding.org/guidelines/v5/content/introduction.html#modelChanges}}  facilitates the linking symbolic events (such as notes) to data related to specific time points in linked audio files. \texttt{<extData>} can contain a standard XML \texttt{<![CDATA[]]>} tag, thus the exact specifications of the linked data are flexible. For the use case in this paper, JSON formatted performance data \cite{crockford06} extracted from an audio file is encoded, following from \cite{devaney19}. Since \texttt{<extData>} must be linked to a symbolic event, some type of transcription of the audio must be available. However, this could range from a fully specified musical score to note data without rhythmic or metrical information. The approach is flexible, as more detailed score information can be added later if it becomes available. 

\section{Encoding Vocal Performance Data With AMPACT}\label{sec:typeset_text}

To demonstrate how performance data can be encoded with \texttt{<extData>}, we provide a set of MEI files of the vocal lines in Rihanna's \emph{Anti} album encoded with note-level performance data extracted using AMPACT. AMPACT\cite{devaney12} is a suite of tools for estimating performance data in monophonic or polyphonic contexts for which a symbolic representation of the musical content is also available. In situations where a full musical score is available, AMPACT uses a score-audio alignment method that identifies distinct onsets and duration for all notes, including notated simultaneities, such as chords, in polyphonic contexts \cite{devaney14}. For audio for which a corresponding score is not available, note-level performance data can be extracted using transcriptions.

\begin{figure}
  \centering
  \includegraphics[width=.75\columnwidth]{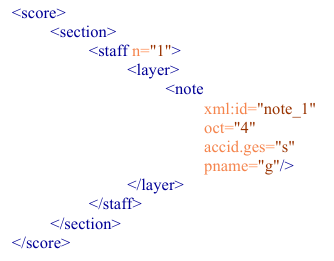}
  \caption{Example of MEI score data encoding. In this example the \texttt{oct} (octave), \texttt{accid.ges} (accidental), and \texttt{pname} (pitch) properties were derived from the note estimates in Tony and the \texttt{xml:id} attribute allows linking with the performance data in Figure \ref{fig:extData}.}
\label{fig:score}
\end{figure}

\subsection{Extracting Performance Data}\label{subsec:extracting}
For each track, we isolated the vocals using Open-Unmix \cite{stoter19} (using a sample rate of 44100 with 0 iterations) and used Tony \cite{mauch15} to estimate the note onsets, duration, and notes in each isolated vocal stem. The Tony note segmentations were hand-corrected, particularly to correctly segment the individual notes in melismas. Note onset, duration, and pitch data were exported from TONY and imported into AMPACT, where it was used as a proxy for score alignment. The annotated note events, with their associated timings, were used as time-frequency regions of importance \cite{devaney17} to estimate the following frame-wise continuous performance descriptors: fundamental frequency (\fo{}), power, spectral centroid, spectral flux, spectral slope, and spectral flatness. Summary descriptors were calculated from the continuous data, using the methods described in \cite{devaney16} to generate four pitch-related descriptors (perceived pitch, jitter, vibrato depth, and vibrato rate), two power-related descriptors (average power and shimmer), and four timbre-related descriptors (average spectral centroid, average spectral flux, average spectral slope, and average spectral flatness). The use of transcription as a proxy for score data in AMPACT facilitated the filtering of any bleed-through in the Unmix-isolated vocal tracks. 

\subsection{Encoding Performance Data}\label{subsec:extracting}
\begin{figure}[h]
  \centering
  \includegraphics[width=.9\columnwidth]{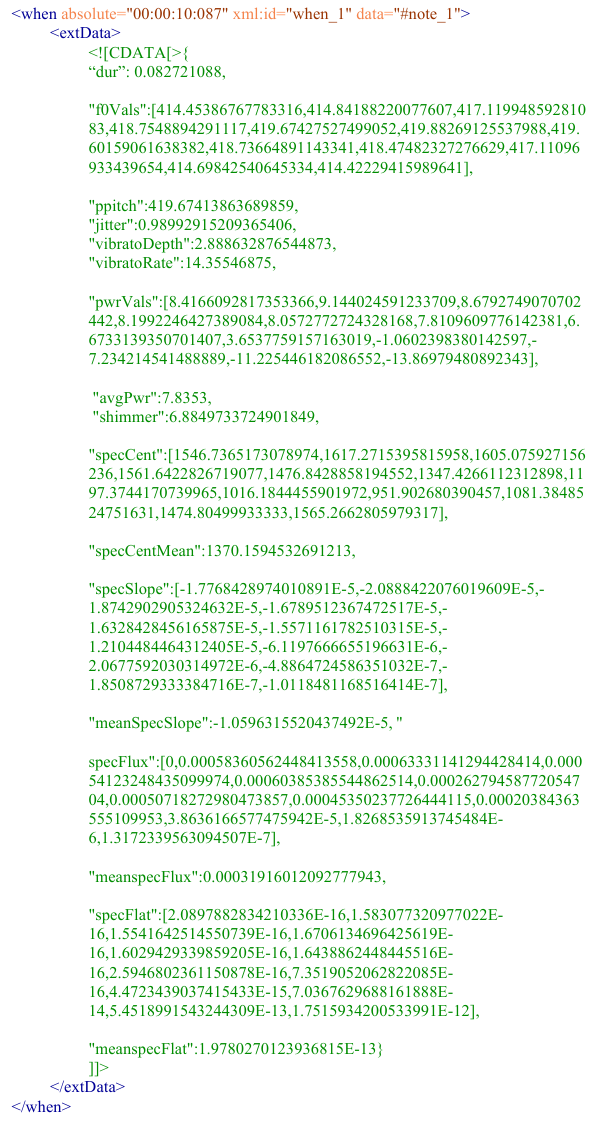}
  \caption{Example of performance data encoded in MEI using the \texttt{<extData>} tag. The performance data is linked to the corresponding note in Figure \ref{fig:score} through the \texttt{data} attribute of the \texttt{<when>} tag. The performance data is encoded as a JSON object, held in a \texttt{<![CDATA[>} tag.} 
\label{fig:extData}
\end{figure}

To generate an encoding, the note data exported from TONY was broken down to the octave, pitch name, and (if present) accidental components required by the MEI format (see Figure \ref{fig:score} for an example of a single note). Metrical information could be included if available, but it is not necessary to produce validated MEI. The continuous and summary performance data estimated with AMPACT for each note were first formatted as a JSON object along with the note duration information exported from Tony. For each note, the JSON object was wrapped in an \texttt{<extData>} tag and linked to the corresponding note with a \texttt{<when>} tag. The start time of each note in the audio file was encoded with the \texttt{absolute} attribute of the \texttt{<when>} tag. See Figure \ref{fig:extData} for an example encoding of the performance data of a single note, which in a full MEI file would be linked to the note in Figure \ref{fig:score}. We used this procedure to encode the vocal tracks in the deluxe edition of Rihanna's \emph{Anti} album.\footnote{\url{https://github.com/jcdevaney/mei-extData/}} 

\section{Conclusions}
Encoding performance data with the \texttt{<extData>} tag in MEI has several advantages. It is flexible in regards to the amount of symbolic data that needs to be encoded and also extensible in terms of adding more detailed symbolic data at a later point in time. The example encoding of the \emph{Anti} demonstrates how it can be used for encoding popular music with under-specified scores. Continuous data can be encoded as easily as summary descriptors, allowing for lower-level data to be easily shared. And finally, as discussed in \cite{devaney20}, it facilitates linking of performance data with a range of other types of data \cite{lewis19,page19,weigl19}, including musical analysis \cite{rizo19,ericson20}, which in turn facilitates research across musical parameters. Our next steps include exemplifying this last point by encoding performance data for the CoSoD dataset \cite{duguay23}.

\newpage

\section{Acknowledgements}
This work was supported by a National Endowment for the Humanities  Digital Humanities Advancement Grant (Grant Number: HAA-281007-21).
\bibliography{ISMIR2023_lbd}

%
%
%
%
%

\end{document}